\documentclass[aps,preprint,superscriptaddress,showpacs,preprintnumbers,amsmath,amssymb]
{revtex4}

\usepackage{ae,graphicx}

\usepackage{epsfig}

\input epsf % somewhere early on in your TeX file

\newcommand{\mb}{$\mu_B$ }

\newcommand{\fr}{$Fe_{1-x}Ru_x$}

\newcommand{\be}{\begin{equation}}
\newcommand{\ee}{\end{equation}}
\newcommand{\ba}{\begin{eqnarray}}
\newcommand{\ea}{\end{eqnarray}}

\begin{document}

\title{Theoretical study of the intrinsic magnetic properties of disordered \fr \ alloys: a
mean-field approach}% Force line breaks

\author{C. Paduani}\email{paduani@fisica.ufsc.br}
\author{N. S. Branco}\email{nsbranco@fisica.ufsc.br}
\affiliation{Departamento de F\'{\i}sica, Universidade Federal de
Santa Catarina, 88040-900, Florian\'{o}polis, SC, Brazil}

\date{\today}

\begin{abstract}
The magnetic properties of the  \fr \ alloy system for 0 $\leq$ x
$\leq$ 0.10 are studied by using a mean-field approximation based
on the Bogoliubov inequality. Ferromagnetic Fe-Fe spin
correlations and antiferromagnetic Fe-Ru and Ru-Ru exchanges have
been considered to describe the temperature dependence of the
Curie temperature and low temperature magnetization. A composition
dependence has been imposed in the exchange couplings, as
indicated by experiments.  From a least-square fitting procedure
to the experimental results an estimation of the interaction
parameters was obtained, which yielded the low temperature
dependence of the magnetization and of the ferromagnetic Curie
temperature. A good agreement was obtained with available
experimental results.

\end{abstract}


\maketitle

\section{Introduction}

    The $Fe$-based alloys with the  4d transition metals have been
intensively investigated since the earliest studies on magnetic
materials. Nevertheless, theoretical and experimental results on
$Fe$-$Ru$ systems are scarce
\cite{clendenen,maurer,knab1,knab2,tian,kobayashi1,kobayashi2,sanchez,saintlager,blachowski,geng}.
Iron and ruthenium are miscible over the entire range of
composition. The iron-rich $Fe$-$Ru$ alloys are ferromagnetic (FM)
at room temperature in the bcc structure\cite{hansen}; the Curie
temperature decreases steadily with the  $Ru$ content. According
to recent investigations in disordered \fr \ alloys, for $x <
0.30$ a single phase is formed with a bcc structure, whereas for
$x \ge 0.30$ there is a crystallographic transition to an hcp
structure \cite{pottker}. In the bcc phase the lattice parameter
has a linear increase with the increase of the $Ru$ concentration.
The experimental results evidenced that antiferromagnetic (AF)
$Fe$-$Ru$ exchanges are settled up in dilute alloys which depends
on the solute concentration.

    First-principles electronic structure calculations on the magnetic phases of iron
compounds in the $CsCl$ structure with 4d elements have  shown
that $FeRu$  indeed has an AF ground state \cite{maurer}. The
introduction of $Ru$ in the immediate neighborhood has been found
to enhance the magnetic moment at $Fe$ sites \cite{kobayashi1}.
Actually, a competition mechanism between FM and AF exchanges is
expected to occur in $Fe$-rich \fr \ alloys, although the FM
$Fe-Fe$ coupling is expected to be overwhelming. Recent
first-principles calculations has also confirmed that with the
introduction of $Ru$ atoms in the bcc iron matrix  the $Fe$ moment
changes appreciably and the average moment decreases steadily
\cite{paduani2}.   The $Ru$ atom as a single impurity in this host
carries a small moment of about 0.27 \mb, which is
ferromagnetically coupled to the surrounding $Fe$ atoms. With the
increase of the distance between $Ru$ atoms larger moments have
been observed  for the $Fe$ atoms in dilute alloys. The contact
hyperfine field has also been found to be very sensitive to the
separation between  $Ru$ atoms in the first shell of neighbors,
and scales with the magnetization.

In this study we apply a mean-field approximation based on the
Bogoliubov inequality  to assess the composition dependence of the
intrinsic magnetic properties of disordered  \fr \ alloys.  Since
these alloys are formed in the bcc structure, mean-field-like
procedures are expected to be a very good approximation to
describe their magnetic behavior. Our model assumes that the
$Fe-Fe$ interaction is ferromagnetic, while $Ru-Ru$ and $Fe-Ru$
interactions are antiferromagnetic. The sites on the lattice are
occupied either by a $Fe$ atom or a $Ru$ atom, according to the
distribution:
\begin{equation}
  {\cal P}(\epsilon_i) = (1-x) \delta(\epsilon_i-1) + x \delta(\epsilon_i), \label{eq:distrib.sitio}
\end{equation}
where $\epsilon_i =1(0)$ for $Fe$($Ru$) atoms. The Hamiltonian reads:
\begin{equation}
   {\cal H} = - \sum_{<i,j>}   J_{ij} S_i S_j,   \label{eq:hamiltonian}
\end{equation}
where the sum runs over all pairs of nearest-neighbor sites and $S_i = \pm 1$,
for all sites $i$. Since the atoms are
randomly distributed in the lattice, the bond between nearest-neighbor
$S_i$ and $S_j$, $J_{ij}$, takes the values $J$ for $Fe-Fe$ pairs, $-\alpha J$ for
$Fe-Ru$ pairs and $-\xi J$ for $Ru-Ru$ pairs, with
probabilities $(1-x)^2$, $2x(1-x)$ and $x^2$, respectively. We
took the assumption  that both $\alpha$ and $\xi$ parameters are
positive. We will show that it is crucial to take into account a
dependence of the exchange interaction on the fraction of $Ru$
atoms. Since from experimental results the lattice parameter
varies with the $Ru$ concentration, this dependence is thereby
expected.

    In the next section we outline the adopted formalism by focusing on the new
features, and in Section \ref{discussion} we present and discuss the results.

\section{Calculational Details}

    The Bogoliubov inequality is a useful way to construct a mean-field-like
approximation to a Hamiltonian ${\cal H}$ which cannot be solved exactly \cite{callen}. It reads:
\begin{equation}
     F({\cal H}) \leq \phi(\zeta) \equiv \left[ F_0 \right] +
 \left[ \left< {\cal H} - {\cal H}_0 \right>_0 \right],    \label{eq:bogoliubov}
\end{equation}
where ${\cal H}_0$ is an exactly solvable tentative Hamiltonian,
$F_0$ is the free energy associated with ${\cal H}_0$, $\left<
\cdots \right>_0$ represents averages made on the ensemble defined
by ${\cal H}_0$ and $\left[ \cdots \right]$ represents the
disorder average. This Hamiltonian depends on the variational
parameter(s) $\zeta$. The right-hand side of the previous equation
is then minimized with respect to this (these) variational
parameter(s), in order to get the best approximation, given the
tentative Hamiltonian ${\cal H}_0$.

    For this work we chose ${\cal H}_0$ to be a combination of single-site and
single-pair Hamiltonians, namely:
\begin{equation}
     {\cal H}_0 = - \gamma_S \sum_{i=1}^{n_1} S_i - \sum_{\{j,k\},j \neq k}^{n_2} J_{ij} S_j S_k
- \gamma_P \sum_{j=1}^{2n_2} S_j,
\end{equation}
where the first sum runs over $n_1$ isolated sites, the second sum
runs over $n_2$ isolated pairs of spins and the last one runs
over the $2 n_2$ sites in the isolated pairs, with $N=n_1+2 n_2$, where
$N$ is the total number of sites.
The variational
parameters are $\gamma_S$ and $\gamma_P$. The configurational
average of the interactions $J_{ij}$ will be made with the
probability distribution:
\begin{eqnarray}
       {\cal P}(J_{ij}) & = & (1-x)^2 \delta(J_{ij}-J) + 2x(1-x) \delta(J_{ij}+\alpha J) \nonumber \\
 & & + x^2 \delta(J_{ij}+\xi J). \label{eq:distrib.ligacao}
\end{eqnarray}
   Note that, if the site occupation is subject to the probability distribution
given by Eq. (\ref{eq:distrib.sitio}), the bonds are no longer
independently distributed since the presence of a $Ru$ atom at a
site forces the eight bonds that emerge from this site to be
either $Ru-Ru$ or $Fe-Ru$. This correlation is \textit{not} taken
into account in Eq. (\ref{eq:distrib.ligacao}). However, since in
our approximation pairs are independent, this correlation is not
present at this level and then we can use Eq.
(\ref{eq:distrib.ligacao}) to make the configurational averages.

    It is easy to show that the free energy associated with the trial
Hamiltonian ${\cal H}_0$ is given by:
\begin{equation}
    F_0 = -kT \ln (Z_S^{N-2n_2} Z_P^{n_2}),
\end{equation}
where $N$ is the number of sites, $k$ is the Boltzmann constant,
$T$ is the temperature, and
\begin{equation}
  Z_S = 2 \cosh (\gamma_S/kT)  \label{eq:ZS}
\end{equation}
and
\begin{equation}
  Z_P(J_{ij}) = 2 \exp(J_{ij}/kT) \cosh(2 \gamma_P /kT)
+ 2 \exp(-J_{ij}/kT) \label{eq:ZP}.
\end{equation}
Therefore:
\begin{equation}
  \left[ F_0 \right] = \int F_0 {\cal P}(J_{ij}) dJ_{ij}.  \label{eq:F0}
\end{equation}
In the same way we obtain:
\begin{eqnarray}
     \left[ \left< {\cal H} - {\cal H}_0 \right>_0 \right] & = &
 -\left( \frac{Nz}{2}-n_2 \right) m^2 \int J_{ij} {\cal P}(J_{ij}) dJ_{ij} + \nonumber \\
  & & (N-2n_2) \gamma_S m + 2 n_2 \gamma_P m,
          \label{eq:HmenosH0}
\end{eqnarray}
where $m$ is the magnetization (see next two equations) and
$z=8$ for the bcc lattice. Then $\phi(\zeta)$ is constructed according to
Eq. (\ref{eq:bogoliubov}).

    The magnetization can be obtained from isolated sites or from isolated
pairs, respectively:
\begin{equation}
  m_S = \left[ \frac{1}{\beta} \frac{\partial \ln Z_S}{\partial \gamma_S}\right] = \tanh(\gamma_S/kT)
        \label{eq:ms}
\end{equation}
and
\begin{eqnarray}
  m_P & = & \left[ \frac{1}{\beta} \frac{\partial \ln Z_P}{\partial \gamma_P}\right] =
2 \sinh(2\gamma_P/kT) \times \nonumber \\
          & &  \left\{ (1-x)^2 \frac{\exp(J/kT)}{Z_P(J)} + \right.
x^2 \frac{\exp(-\xi J/kT)}{Z_P(-\xi J)} + \nonumber \\
          & & \left. 2x(1-x) \frac{\exp(-\alpha J/kT)}{Z_P(-\alpha J)} \right\},
         \label{eq:mp}
\end{eqnarray}
where $\beta=1/kT$.

    Minimizing the approximated free energy with respect to $\gamma_S$
and taking into account the above expressions for the magnetization, we obtain:
\begin{equation}
      \gamma_S = \frac{z}{z-1} \gamma_P.  \label{eq:relacaoentreosgamas}
\end{equation}

    We have chosen $n_2=z N/2$, which is the maximum number of pairs for
a lattice of $N$ sites and coordination number $z$. Also, $\phi(\zeta)$ decreases
when $n_2$ increases and, therefore, the value we chose for $n_2$ leads to
the minimum value physically meaningful for $\phi(\zeta)$. We believe this to lead
to the best approximation possible for the true free energy within our procedure.

    By imposing that the two expressions for the magnetization, i.e.,  Eqs. (\ref{eq:ms}) and
(\ref{eq:mp}) are equal, expanding them for small $\gamma_S$
and $\gamma_P$, and using Eq. (\ref{eq:relacaoentreosgamas}), we
obtain:
\begin{eqnarray}
   \frac{z}{2(z-1)} & = & \left\{ \frac{(1-x)^2}{1+\exp(-2J/kT)} + \frac{2x(1-x)}{1+\exp(2\alpha J/kT)} +
\right. \nonumber \\
  & & \left. \frac{x^2}{1+\exp(2\xi J/kT)} \right\}. \label{eq:Tc}
\end{eqnarray}

    This expression with $z=8$ can be used to obtain the critical temperature for
the bcc lattice as a function of $x$. The experimental values
of these critical temperatures were reported in Ref.
\onlinecite{pottker}. We have made a best fitting procedure in
order to evaluate the parameters $\alpha$ and $\xi$; details will
be given and the results discussed in Section
\ref{discussion}. Note that, since we have made an expansion for small
$\gamma_S$ and $\gamma_P$, the previous expression is valid only near
$T_c$.

    We can also evaluate the magnetization, again imposing that $m_S=m_P$
(see Eqs. \ref{eq:ms} and \ref{eq:mp})
and solving it for $\gamma_S$ with the help of Eq.
(\ref{eq:relacaoentreosgamas}). Therefore the value of $\gamma_S$
can be used in Eq. (\ref{eq:ms}) to evaluate $m_S$. See next
section for results and discussion.

\section{Results and Discussion} \label{discussion}

    The procedure outlined in the previous section can be used to obtain the
value of the exchange constant, $J$, for pure iron. In this case,
the experimental value for the critical temperature is $T_c=1043$
$K$; from Eq. (\ref{eq:Tc}) with $x=0$, we obtain $J=12.9$
$meV$. This value agrees with that one found in Ref.
\onlinecite{pottker} and is within the range $10-50$ $meV$, as
expected for $Fe$, $Co$, and $Ni$ \cite{kaul}.

    Eq. (\ref{eq:Tc}) can also be used to adjust the parameters to fit the
experimental values for the critical temperature as function of
the $Ru$ fraction, $x$ (see Table \ref{tab:Tcxx}). The
experimental values were taken from Ref. \onlinecite{pottker}. To
show that it is indeed necessary to take into account a variation
of the AF interaction constants with $x$, we have plotted in Fig.
\ref{fig:Tcxx} the critical temperatures given by Eq.
(\ref{eq:Tc}) with $\alpha=\xi=1.0$ (squares) and
$\alpha=\xi=0.79$ (triangles). This last value is the one which
makes the experimental and theoretical values coincide for $x=0$
and $x=0.02$. Clearly, a constant AF interaction will not adjust
the experimental values. We then propose a concentration
dependence for the AF interactions, as has been pointed out in
Ref. \onlinecite{pottker}. Since we have only five experimental
values of $T_C$  for the disordered alloy, we will assume that
(see Eq. (\ref{eq:distrib.ligacao})):
\begin{equation}
   \alpha \equiv \xi = \alpha_{0} - \alpha_{1} \; x. \label{eq:parameters}
\end{equation}
The values we obtain with a non-linear least-square-fitting method
are:
\begin{equation}
   \alpha_{0} = 0.54(2); \;\; \alpha_{1} = 5.4(4), \label{eq:adjustedparameters}
\end{equation}
where the values in parentheses are the errors in the last decimal
figure. In Fig. \ref{fig:Tcxx} the theoretical curve is
represented by a dashed line, while the experimental results are
represented by open circles (error bars are smaller than the
points). As it can be seen, the agreement between the adjusted
curve and the experimental is excellent.

\begin{table}[b]
\caption{Critical temperatures for \fr; figures in parenthesis are
the errors and apply to the last figure (values taken from Ref.
\onlinecite{pottker}).} \label{tab:Tcxx}
\renewcommand{\tabcolsep}{.5cm}
\begin{tabular}{cc}
\hline \hline
$x$ & $T_c$  \\
\hline
0.0 & 1043 \\
0.02 & 968(2)  \\
0.04 & 928(2)  \\
0.06 & 908(2)  \\
0.10 & 838(2) \\
\hline \hline
\end{tabular}
\end{table}

    We have also calculated the magnetization for some values of $x$,
as outlined at the end of the previous section. The results are
depicted in Fig. \ref{fig:magnetizacao}: as expected, the critical
temperature decreases as the concentration of $Ru$ is increased.
Since we have used a mean-field approximation, static critical
exponents assume their classical values. Therefore, the question
of universality classes cannot be addressed by the present
procedure. We are now performing a Monte Carlo simulation on this
alloy to calculate thermodynamic quantities and some critical
exponents. Note the inset  in Fig. \ref{fig:magnetizacao}: we
expect that the zero-temperature value of the magnetization varies
with $x$, since the introduction of AF interactions will freeze
some of the spins in the reversed position, when compared to the
$Fe$ background. In fact, $m(T=0)$ decreases as the fraction of
$Ru$ is increased, as noted for $x=0.02, 0.04$ and $0.06$. For
$x=0.10$ the AF bonds are no more present: for the values of the
adjusted parameters $\alpha_{0}$ and $\alpha_{1}$ (see Eqs.
(\ref{eq:parameters}) and (\ref{eq:adjustedparameters})) and for
$x=0.10$, $\alpha=\xi=0$ and the $Ru$ atoms act as nonmagnetic
impurities. Since the fraction of magnetic ($Fe$) atoms for
$x=0.10$ is well above the percolation threshold for the bcc
lattice, we expect that nearly all $Fe$ atoms take part in the
infinite cluster; therefore, the value of the magnetization at
$T=0$, for $x=0.10$, should be close to $1$. As the temperature is
raised from zero, the AF bonds (which are weaker than the FM ones)
disorder for small values of $T$ and the magnetization increases.
Nevertheless, for finite (but still low) temperatures, the
behavior of the magnetization is not monotonic with respect to
$x$. This result is a consequence of the balance between two
effects: as $x$ increases, $m$ decreases due to a greater number
of AF bonds but increases due to the weakening of these bonds.
This feature explains the behavior seen in the inset of Fig.
\ref{fig:magnetizacao}. The fact that the magnetization returns to
1 as the temperature is increased, for $x=0.02, 0.04$ and $0.06$,
may be an artifact of the mean-field approximation: this aspect
will be clarified by the Monte Carlo simulation.

    In summary, we have calculated the interaction constants for the \fr \
system by using a mean-field approximation based on the Bogoliubov
inequality. The agreement between our theoretical predictions and
the results of experiments is excellent and shows that it is
necessary to take into account a concentration dependence of the
antiferromagnetic interaction strength. We have also calculated
the magnetization as a function of the temperature for some $x$
values, and discussed in detail the expected low temperature
behavior. At $T=0$ the magnetization $m$ decreases as the $Ru$
content $x$ increases for $0 \leq x < 0.10$ but attains the value
$1.0$ for $x = 0$ and $0.10$ . At low but still finite
temperatures the dependence of $m$ on $x$ is nonmonotonic, owing
to a competition mechanism which arises from the effects
introduced as $Ru$ atoms substitutes for $Fe$: the appearance of
antiferromagnetic interactions and their weakening due to the
dependence of the antiferromagnetic constant exchange on $x$. We
are now doing a Monte Carlo simulation on this system to calculate
thermodynamic quantities and critical exponents.

\begin{acknowledgments}
The authors would like to thank the Brazilian Agencies FAPESC, CNPq,
and CAPES for partial financial support and I. J. L. Diaz and T. P.
de Oliveira for a critical reading of the manuscript.
\end{acknowledgments}

\bibliographystyle{unsrt}
\bibliography{references}
%\bibliographystyle{plain}:Entries are ordered alphabetically;
% \bibliographystyle{unsrt} Entries are not ordered alphabetically, but in the order they are
%first referenced.
%\bibitem{han}  M. Hansen and K. Ardenko, {\em Constitution of Binary
%Alloys} (McGraw-Hill, New York, 1958).

\newpage

\begin{figure}[h]
\begin{center}
\leavevmode
%\epsffile{Tcxx.eps}
\caption{Critical temperature as a function of the $Ru$
concentration. Experimental points are represented by circles.
Squares (triangles) represent constant AF interactions, $J_{AF}$,
with $J_{AF}/J = 1.0(0.79)$, where $J$ is the ferromagnetic
interaction strength between $Fe$ atoms. The dashed line
represents results from a least-square fitting by taking into
account that $J_{AF}$ varies with $x$ in the form $J_{AF} = 0.54 -
5.4 \; x$.}
\label{fig:Tcxx}  %Figura 1
\end{center}
\end{figure}

\begin{figure}[h]
\begin{center}
\leavevmode
%\epsffile{magnetizacao.eps}
\caption{Magnetization versus temperature for five different $x$
values. The inset shows details of the magnetization behavior at
low temperatures. The legends are the same for both graphs; the
outermost curve corresponds to $x=0$ while the innermost is for
$x=0.10$.}
\label{fig:magnetizacao}  %Figura 2
\end{center}
\end{figure}

\end{document}